\documentclass[conference]{IEEEtran}

\usepackage{graphicx}
\usepackage{multirow}
\usepackage{array}
\usepackage{subfigure}
\usepackage{amsmath}

\begin{document}
\title{TSEP: Threshold-sensitive Stable\\ Election Protocol for WSNs}

\author{\IEEEauthorblockN{A. Kashaf, N. Javaid, Z. A. Khan$^{\S}$, I. A. Khan$^{\ddag}$\\}
        Dept. of Electrical Engineering, COMSATS, Islamabad, Pakistan. \\
        $^{\S}$Faculty of Engineering, Dalhousie University, Halifax, Canada.\\
        $^{\ddag}$Dept. of Computer Science, COMSATS, Abbottabad, Pakistan.

        }

\maketitle
\begin{abstract}

Wireless Sensor Networks (WSNs) are expected to find wide applicability and increasing deployment in near future. In this paper, we propose a new protocol, Threshold Sensitive Stable Election Protocol (TSEP), which is reactive protocol using three levels of heterogeneity. Reactive networks, as opposed to proactive networks, respond immediately to changes in relevant parameters of interest. We evaluate performance of our protocol for a simple temperature sensing application and compare results of protocol with some other protocols LEACH, DEEC, SEP, ESEP and TEEN. And from simulation results it is observed that protocol outperforms concerning life time of sensing nodes used.
\end{abstract}

\section{Introduction}
Advancements in technology leading to a move from wired to wireless domain. Functionality of wireless devices is dependent upon their battery life time. Wireless sensors are small, low power devices deployed in a field in large number. These sensing nodes have many uses like monitoring physical or environmental conditions, such as temperature, humidity, sound, motion, etc. Wireless Sensor Networks (WSNs) enable us to use these small sensor nodes for multiple applications like military applications; manufacturing, end user applications, area monitoring, waste water monitoring, etc.

In WSNs, nodes sense data and send information to sink. Wireless sensor nodes can be mobile or stationary and can be deployed in their environment randomly or with a proper deployment mechanism. For random deployment there is even distribution of nodes over the field, while for regular deployment nodes are static. Some of energy of nodes is consumed during sensing as well as some part of it is reduced due to transmission and reception of data. Practically, it is not possible to replace or recharge batteries of nodes once deployed. WSN must operate without human involvement so the main focus is to increase network life in any way and for this purpose many protocols are introduced. Routing protocols can be classified on the basis of their applications into following two categories:\\

\textit{a. Proactive Routing Protocols:}
Nodes in network provide a continuous report of data, nodes keep on sensing, turn on their transmitters and transmit, so suitable for applications where information on regular basis is required.

\textit{b. Reactive Routing Protocols:}
Nodes sense data continuously however, transmit only at the time when there is a drastic change in sensed value, so, reactive networks are suitable for time critical applications.

In routing protocols clustering reduces energy consumption in sensor nodes [1, 4, 10, 11]. When clusters are formed, election of CHs can be done on the basis of energy of nodes or on probability of nodes to be elected as CHs. After clusters formation each node transmits data during its time slot and as the last node transmits data, schedule is repeated. The total time spent in completing this schedule is called frame time.

Direct Transmission, a traditional approach in which each node senses data, turns on its transmitter and sends its data directly to sink. For nodes placed closed to sink, data transmission causes less reduction in energy however for nodes at far distances from sink will die more quickly [1].

%The remaining paper is organized in such a way that section II provides related work and motivation, section III is about proposed protocol. Section VI presents performance analysis and simulation results. Section V provides conclusion according to obtained results.

\begin{figure}
\begin{center}
\includegraphics[height=2in,width=2.5in,angle=0]{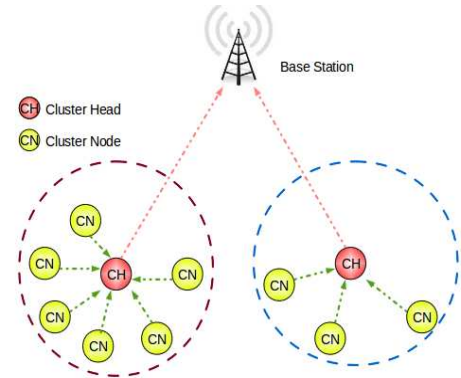}
\caption{\small \sl Cluster formation in WSN.\label{fig:Stupendous}}
\end{center}
\end{figure}

\section{Background}
Clustering procedures are engaged in dealing with energy control. Description of some of concerned protocols is provided in this section.

In Minimum Transmission Energy (MTE) [2], transmission is done through the paths where minimum transmission power is depleted. So, under MTE, nodes that are at large distances from the sink will die later, while nodes near sink act as conveys and so will die sooner.

\subsection {Low Energy Adaptive Clustering Hierarchy (LEACH)}

 LEACH is a proactive routing protocol proposed in [3]. In a network hundreds and thousands of nodes dispersed randomly for even distribution of load among nodes. These nodes sense data, transmit it to their associated CHs which receive, aggregate and then convey this data to the sink or to the Base Station (BS). All the nodes deployed in field are homogeneous and constrained in energy. To divide burden among nodes, improve network life clusters are formed. Nodes are made to become CHs on turns. Nodes randomly elect themselves as CHs and it is done in a way that each node becomes CH once in an epoch $1 /P$ [1, 7, 8, 9]. CHs selection is done on probabilistic basis, each node generates a random number $r$ inclusive of 0 and 1, if generated value is less than threshold computed by formula given below, then this node becomes CH.

\begin{eqnarray}
T_{n} = \left\{ \begin{array}{rl}
 \frac{p}{1-p[r. mod \frac{1}{p}]} &\mbox{ if $n \epsilon  G'$} \\
  0 &\mbox{ otherwise}
       \end{array} \right.
\end{eqnarray}

 After clusters formation, each CH broadcasts a TDMA schedule for nodes associated with it. Nodes sense and transmit data to associated CHs during time slots assigned to them. Once each node in a cluster sent data, frame is repeated. In WSNs, main purpose is to control energy consumption and hence to increase network life. LEACH is not useful to be used in large areas due to energy constraint. In LEACH once attributes are selected, they can not be changed.

\subsection{Stable Election Protocol (SEP)}

As described in [4], heterogeneity is introduced in SEP protocol, which is based on two levels of heterogeneity. A fraction $m$ of total $n$ nodes is provided with an additional energy factor $\alpha$, which are called advanced nodes. So, probabilities of normal nodes and advanced nodes to become CHs are $p_{nrm}=\frac{p_{opt}}{1+m.\alpha}$ and $p_{adv}=\frac{p_{opt}.(1+\alpha)}{1+m.\alpha}$ respectively, where $p_{opt}$ is the optimal probability of each node to become CH. CHs election in SEP is done randomly on the basis of probability of each type of node as in LEACH. Nodes sense data and transmit it to associated CH which convey it to BS. By increasing $m$ or $p_{adv}$, we can further improve our system. So, SEP results in increased stability period and network life due to advance nodes however two level heterogeneity also caused increased throughput.

\subsection{Enhanced Stable Election Protocol (ESEP)}

An extension of SEP proposed in [5] considers three types of nodes, normal nodes, intermediate nodes and advance nodes. Where, advance nodes are in a fraction of total nodes with an additional energy as in SEP and a fraction of nodes with some extra energy greater than normal nodes and less than advance nodes, called intermediate nodes, while rest of the nodes are normal nodes. As in SEP, in ESEP CHs are selected depending on probability of each type of node. However, energy dissipation is controlled to some extent due to three levels of heterogeneity.

\subsection{Threshold Sensitive Energy Efficient sensor Network protocol (TEEN)}

A reactive routing network protocol, used for time critical applications. In TEEN transmission is done only when a severe change occurs in field. It is a threshold sensitive protocol based on two threshold levels, hard threshold and soft threshold. Whenever the sensed attribute's value becomes equal or greater than hard threshold, nodes turn on their transmitters and data is conveyed to CHs. And for the second time they transmit only in case when the difference between sensed value and previously saved value at which transmission was done is greater than or equal to soft threshold. So, energy consumption as well as throughput is reduced, hence network life and stability period are improved than other protocols discussed above [6].\\

In SEP, ESEP, TEEN, at cluster selection time, attributes can be changed according to requirements. SEP and ESEP being heterogeneity aware protocols improve stability period and network life time but here a drawback of heterogeneity is this that throughput is also increased which is eventually causing decrease in network life time. To control trade off between energy efficiency, accuracy and response time dynamically our proposed protocol is observed to be better than other protocols discussed above.

\section{Proposed Protocol}
In this section we describe our new protocol TSEP (Threshold sensitive Stable Election Protocol) which has two main features: ``It is reactive routing protocol'', as transmission consumes more energy than sensing and it is done only when a specific threshold is reached and ``Three levels of heterogeneity'' .

To describe whole protocol clearly we particularly discuss about energy model and how optimal number of clusters can be computed.
For three levels of heterogeneity, nodes with different energy levels are:

\begin{enumerate}
  \item Normal Nodes
  \item Intermediate Nodes
  \item Advance Nodes\\
\end{enumerate}

Advance nodes having energy greater than all other nodes, intermediate nodes with energy in between normal and advance nodes while remaining nodes are normal nodes. Intermediate nodes can be chosen by using $b$, a fraction of nodes which are intermediate nodes and using the relation that energy of normal nodes is $\mu$ times more than that of normal nodes. In SEP energy for normal nodes is $ E_o$, for advance nodes it is $E_{ADV}=E_o(1+\alpha)$ and energy for intermediate nodes can be computed as $ E_{INT}= E_o(1+\mu)$, where $ \mu =\frac{\alpha}{2}$.

So total energy of normal nodes, advance nodes and for intermediate nodes will be, $n.b(1+{\alpha})$, $ nE_o.(1-m-bn)$, and $n.m.E_o.(1+\alpha)$ respectively.

So, the total Energy of all the nodes will be, $nE_o.(1-m-bn)+n.m.E_o.(1+ \alpha)+ n.b.(1+\mu)=n.E_o(1+ m\alpha + b\mu)$.

Where, $n$ is number of nodes $m$ is proportion of advanced nodes to total number of nodes $n$ with energy more than rest of nodes and $b$ is proportion of intermediate nodes.

The optimal probability of nodes, which are divided on the basis of energy, to be elected as a CH can be calculated by using following formulas:

\begin{equation}
p_{nrm}=\frac{p_{opt}}{1+m.\alpha+b.\mu}
\end{equation}

\begin{equation}
p_{int}=\frac{p_{opt}.(1+\mu)}{1+m.\alpha+b.\mu}
\end{equation}

\begin{equation}
p_{adv}=\frac{p_{opt}.(1+\alpha)}{1+m.\alpha+b.\mu}
\end{equation}

Now to ensure that CH selection is done in the same way as we have assumed, we have taken another parameter into consideration, which is threshold level. Each node generates randomly a number inclusive of 0 and 1, if generated value is less than threshold then this node becomes CH [1], [12]. For all these type of nodes we have different formulas for the calculation of threshold depending on their probabilities, which are given below:

\begin{eqnarray}
T_{nrm} = \left\{ \begin{array}{rl}
 \frac{p_{nrm}}{1-p_{nrm}[r. mod \frac{1}{p_{nrm}}]} &\mbox{ if $n_{nrm} \epsilon  G'$} \\
  0 &\mbox{ otherwise}
       \end{array} \right.
\end{eqnarray}

\begin{eqnarray}
T_{int} = \left\{ \begin{array}{rl}
 \frac{p_{int}}{1-p_{int}[r. mod \frac{1}{p_{int}}]} &\mbox{ if $n_{int} \epsilon  G''$} \\
  0 &\mbox{ otherwise}
       \end{array} \right.
\end{eqnarray}

\begin{eqnarray}
T_{adv} = \left\{ \begin{array}{rl}
 \frac{p_{adj}}{1-p_{adv}[r. mod \frac{1}{p_{adv}}]} &\mbox{ if $n_{adv} \epsilon  G'''$} \\
  0 &\mbox{ otherwise}
       \end{array} \right.
\end{eqnarray}
$G'$, $G''$ and $G'''$ are the set of normal nodes, intermediate nodes and set of advanced nodes that has not become CHs in the past respectively, so ensuring that the equations (2), (3) and (4) are working.\\

Average total number of CHs per round will be:

\begin{equation}
n.(1-m-b).p_{nrm}+n.b.p_{int}+n.m.p_{adj}=n.p_{opt}
\end{equation}
Although, average number of CHs is same as that of LEACH, SEP and ESEP. However, here a good aspect of TSEP is energy dissipation is reduced due to energy heterogeneity.

At the start of each round, here takes place the phenomenon of cluster change. In case of TSEP, at cluster change time, the CH broadcasts the following parameters\\
\begin{itemize}
  \item \textbf{Report Time (TR):} Time period during which reports are being sent by each node successively\\
  \item \textbf{Attributes(A):} The physical parameters about which information is being sent.\\
  \item \textbf{Hard Threshold (HT):} An absolute value of sensed attribute beyond which node will transmit data to CH. As if sensed value becomes equal to or greater than this threshold value, node turns on its transmitter and sends that information to CH.\\
  \item \textbf{Soft Threshold (ST):} The smallest sensed value at which the nodes switch on their transmitters and transmit.\\

\end{itemize}

All nodes keep on sensing environment continuously. As parameters from attribute set reaches hard threshold value, transmitter is turned on and data is transmitted to CH, however this is for the first time when this condition is met. This sensed value is stored in an internal variable in the node, called Sensed Value (SV). Then for second time and the other, nodes will transmit data if and only if sensed value is greater than hard threshold value or if difference between currently sensed value and the value stored in SV variable is equal to or greater than soft threshold. So, by keeping these both thresholds in consideration, number of data transmissions can be reduced, as transmission will only take place when sensed value reaches hard threshold. And further transmissions are lessened by soft threshold, as it will eliminate transmissions when there is a small change in value, even smaller than interest. Some of important features are described below:\\
\begin{enumerate}
  \item Time critical data reaches the user almost instantaneously.
  \item Nodes keep on sensing continuously but transmission is not done frequently, so energy consumption is much more less than that of proactive networks.
  \item At time of cluster change, values of soft threshold, TR and A are transmitted afresh and so, user can decide how often to sense and what parameters to be sensed according to the criticality of sensed attribute and application.
  \item The user can change the attributes depending on requirement, as attributes are broadcasted at the cluster change time.
\end{enumerate}
	
One of the main trades off of this scheme is that if threshold is not reached, user will not get any information from network and even if one or all the nodes die, system will not come to know about that. So, it is not useful for those types of applications where a data is required continuously.\\

\section{simulations And Discussions}
 For performance evaluation we used MATLAB. Our goals in doing simulations was to compare performance of TSEP with SEP, ESEP, LEACH, and TEEN protocols on the basis of energy dissipation and longevity of network.

Performance metrics used in the simulations are:\\
\begin{enumerate}
  \item Stability period, the period from the start of the network operation and the first dead node.
  \item Instability period, the period between the first dead node and last dead node.
  \item Number of alive nodes per round.
  \item Number of dead nodes per round.
  \item Throughput, number of packets sent from cluster heads to base station.\\
\end{enumerate}
	
A network consisting of 100 nodes, placed randomly in a region of MxM and a BS located in the center is considered. We performed simulations for different values of $\alpha$ and $m$ while keeping $b$ constant that is 0.3. For the first case $\alpha= 1$,$ m=0.1$ , for second case $\alpha= 3$ and $ m=0.2$. This is done to observe change in network's stability, life and throughput relative to increase in number of advance nodes and their energies. Since $p_{opt} = 0.1$, is the optimal probability of CHs, by using equations (2), (3) and (4) we obtained different probabilities for each type of nodes in accordance with different values of $\alpha$ and $m$. Other parameters used in simulations are shown in Table 1.

By using equations (5), (6) and (7), CHs election for normal, intermediate and advance nodes respectively, can be known.

\begin{table}[ht]
\centering
\begin{tabular}{|c |c |} % centered columns (4 columns)
\multicolumn{2}{c}{Table 1. Parameter Settings}\\
\hline
\textbf{ Parameters} & \textbf{Value}  \\
\hline
$ E_{elect}$ & 50nJ/bit\\
%heading
\hline                  % inserts single horizontal line
  $E_{DA}$ & 5nJ/bit/message  \\
\hline
 % inserting body of the table
 $\epsilon_{fs} $& 10pJ/bit/$m^2$  \\
\hline
 $\epsilon_{mp} $& 0.0013pJ/bit/$m^4$    \\
\hline
 $E_o$ & 0.5J \\
 \hline
 $K$ & 4000 \\
 \hline
 $P_{opt}$ & 0.1 \\
\hline
 $n$ & 100\\
 \hline
 $\alpha$ & 1 \\
 \hline
 $m$ & 0.1 \\\hline
\end{tabular}
\end{table}

\begin{figure}
\begin{center}
\includegraphics[height=1.8in,width=3in,angle=0]{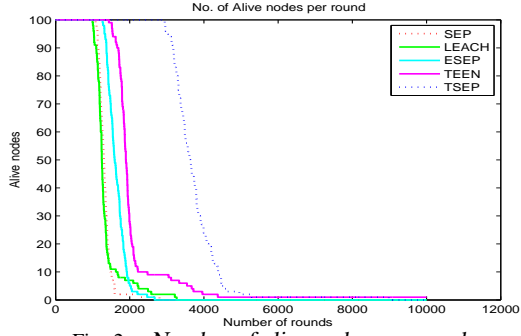}
\vspace{-0.5cm}
\caption{\small \sl Number of alive nodes per round.\label{fig:Stupendous}}
\end{center}
\end{figure}

\begin{figure}
\begin{center}
\includegraphics[height=1.8in,width=3in,angle=0]{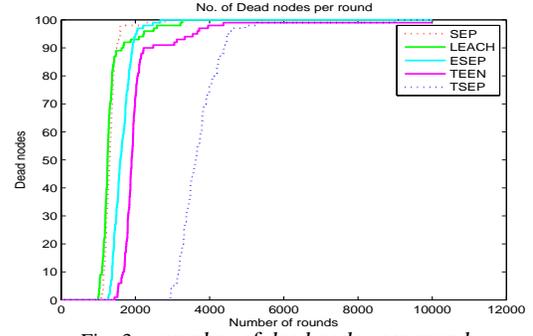}
\vspace{-0.5cm}
\caption{\small \sl number of dead nodes per round.\label{fig:Stupendous}}
\end{center}
\end{figure}
\begin{figure}
\begin{center}
\includegraphics[height=1.8in,width=3in,angle=0]{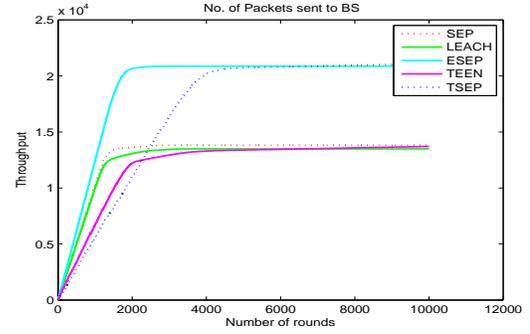}
\caption{\small \sl Number of packets sent from CHs to BS.\label{fig:Stupendous}}
\end{center}
\end{figure}

\begin{figure}[htb]
  \centering
\subfigure[]{\includegraphics[height=5 cm,width=9 cm]{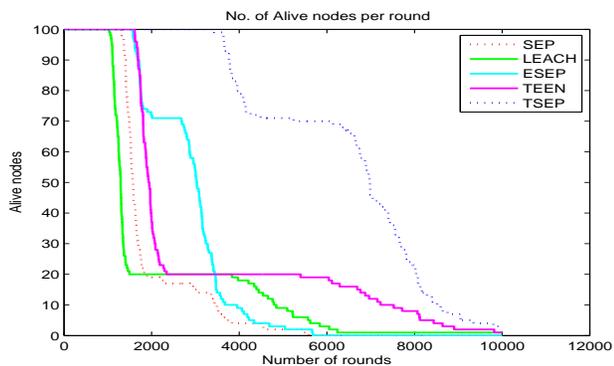}}
\subfigure[]{\includegraphics[height=5 cm,width=9 cm]{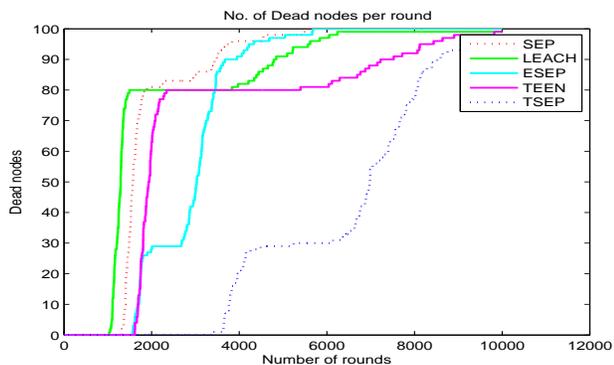}}
\subfigure[]{\includegraphics[height=5 cm,width=9 cm]{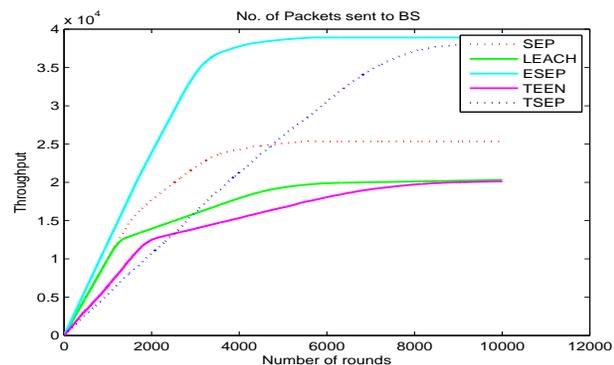}}
\caption{Shows LEACH, SEP, ESEP, TEEN, TSEP behavior in the presence of heterogeneity with $\alpha$ = 3 and m = 0.2 (a) Alive nodes per round; (b) Dead nodes per round; (c) Packets sent from CHs to BS}
\end{figure}

Fig.2 and Fig.3 show comparison of protocols SEP, LEACH, TEEN, ESEP and TSEP regarding alive and dead nodes, relative to number of rounds. Comparing all these protocols, SEP and LEACH being heterogeneous, probability based protocols result in approximately equal stability period and network life. As in SEP and LEACH, CHs selection is based on probability, while, if LEACH would be considered with homogeneity then there would be a large difference. ESEP with three levels of heterogeneity and probability based protocol obviously shows better results than SEP and LEACH, as can be concluded through equations (1-3). Due to availability of more nodes with extra energy ESEP results in increased stability period than SEP and LEACH. It is observed that in TEEN, stability period is greater than all other protocols discussed. As it is threshold based protocol and here transmission is done at only some certain conditions. Nodes keep on sensing and so energy consumption is less than other protocols resulting in increased stability period and network life. The newly proposed protocol TSEP also being threshold based protocol with an additional feature of three levels of heterogeneity results in increased stability period and network life even greater than that of TEEN.

Fig.4 shows throughput that is data sent from CHs to the BS. TSEP and TEEN being threshold sensitive protocols, show better results than all others, as here transmission rate is less so energy consumption will be less than others but due to three level heterogeneity throughput in ESEP and TSEP is greater than SEP, LEACH and TEEN. Comparing TSEP and ESEP, both are having three levels of heterogeneity, but due to threshold sensitivity of TSEP throughput in TEP is less than ESEP.

From Fig.2, Fig.3 and Fig.4 it can be clearly concluded that stability period and network life time are greater in TSEP, than all other protocols. Nodes tend to die slowly in TSEP, as in TSEP a major part of energy is consumed in sensing; while transmission of data is done only at conditions when hard threshold value is achieved by sensed node or is exceeded.

Same is the case for graphs Fig.5(a), (b) and(c), where $\alpha$ = 3 and m = 0.2. In this case, energy of nodes as well as probability of CHs is also increased. So, more number of nodes will be available with extra energy. As shown in Fig.5(a), Fig.5(b), by increasing number of advanced nodes and $\alpha$ stability period and network life are increased.
By Fig. 5(c) it can be seen that throughput, that is data sent from CHs to BS is also increased. It happens because of three level heterogeneity. So, it can be clearly seen that there are noticeable differences among the protocols in accordance with alive nodes, dead nodes and throughput.

By performing simulations in MATLAB, it is observed that:\\
\begin{itemize}
  \item TSEP has enhanced stability period than all other protocols. This is shown in Fig.2, Fig.3 and Fig.5(a), Fig.5(b).
  \item	The network life for TSEP was increased as compared to others.
  \item Increase and decrease in number of alive and dead nodes respectively.
  \item Increased throughput due to three level heterogeneity and decrease in throughput due to threshold sensitivity as can be observed in Fig.3 and Fig.5(c).\\
\end{itemize}
	
\section{conclusions}
In this paper TSEP, reactive routing protocol is proposed where nodes with three different levels of energies. CHs selection is threshold based, due to three levels of heterogeneity and being reactive routing network protocol, it causes increase in stability period and network life. In comparison with SEP, LEACH, ESEP and TEEN it can be concluded that our protocol will perform well in small as well as large sized networks.

\end{document}